\documentclass[12pt]{article}
\usepackage{amsmath}
\usepackage{psfrag,epsf}
\usepackage[pdftex]{graphicx}
\usepackage{enumerate}
\usepackage{natbib}
\usepackage{url} 

\usepackage{graphicx}  
\usepackage{setspace}  
\usepackage{textcomp} 
\usepackage{indentfirst} 
\usepackage{booktabs,array,arydshln} 
\usepackage[T1]{fontenc} 
\usepackage[utf8]{inputenc} 
\usepackage{newtxmath} 

\usepackage{threeparttable}

\usepackage{bbm}
\usepackage{bm}
\usepackage{multirow}
\usepackage{caption}
\usepackage{subcaption}

\newcommand{\blind}{0}

\begin{document}

\def\spacingset#1{\renewcommand{\baselinestretch}%
{#1}\small\normalsize} \spacingset{1}


\if0\blind
{
  \title{\bf Forbidden Knowledge and Specialized Training: A Versatile Solution for the Two Main Sources of Overfitting in Linear Regression}
  \author{Chris Rohlfs\\\hspace{.2cm}\\
    Department of Electrical Engineering, Columbia University}
  \maketitle
} \fi

\if1\blind
{
  \bigskip
  \bigskip
  \bigskip
  \begin{center}
    {\LARGE\bf Forbidden Knowledge and Specialized Training: A Versatile Solution for the Two Main Sources of Overfitting in Linear Regression}
\end{center}
  \medskip
} \fi

\bigskip
\begin{abstract}
Overfitting in linear regression is broken down into two main causes. First, the formula for the estimator includes `forbidden knowledge' about training observations' residuals, and it loses this advantage when deployed out-of-sample. Second, the estimator has `specialized training' that makes it particularly capable of explaining movements in the predictors that are idiosyncratic to the training sample. An out-of-sample counterpart is introduced to the popular `leverage' measure of training observations' importance. A new method is proposed to forecast out-of-sample fit at the time of deployment, when the values for the predictors are known but the true outcome variable is not. In Monte Carlo simulations and in an empirical application using MRI brain scans, the proposed estimator performs comparably to Predicted Residual Error Sum of Squares (PRESS) for the average out-of-sample case and unlike PRESS, also performs consistently across different test samples, even those that differ substantially from the training set.
\end{abstract}

\noindent%
\vfill

\newpage
\spacingset{1.45} 
\section{Introduction} \label{Overfitting introduction}

Empirical models are known to forecast more accurately among \emph{in-sample} observations used for fitting than on previously unseen \emph{out-of-sample} observations. The difference is greatest for complex models with large numbers of estimated parameters; thus, forecasting approaches that are selected based on in-sample performance are excessively complicated and therefore especially disappointing out-of-sample. This process of using in-sample criteria and arriving at models that contain too many parameters is known as \emph{overfitting}.

The widely accepted remedy for overfitting is to split one's dataset into a \emph{training sample} for parameter estimation, a \emph{validation sample} to evaluate broader modeling decisions, and a \emph{test sample} to assess the performance of the final selected model \citep{HastieTibshiraniFriedman2009}. Related to this use of holdout samples is leave-one-out or ``jackknife’’ cross-validation, through which the fit for each case $i$ is determined by separately re-running the model on the other $n-1$ cases; \citet{ArlotCelisse2010} provide a useful survey of this and other cross-validation approaches. The policy of splitting data into training, validation, and test sets addresses many pernicious forms of overfitting and is generally regarded as effective. Nevertheless, it substantially limits the numbers of observations available for any of the three tasks of training, validation, or testing. The jackknife approach avoids the loss in sample sizes but greatly increases the computational burden of assessing model accuracy.

In a linear regression context, the predominant measures of fit, $R^2$ and adjusted $R^2$ \citep{Theil1961} suffer from overfitting, but the well-known Predicted Residual Error Sum of Squares (PRESS) by \citet{Allen1971, Allen1974} solves the issue by exploiting features of the least squares formula that enable jackknife validation with minimal computation. But while PRESS corrects overfitting for the average out-of-sample case, it does not differentiate among test cases or accurately project models' performance on unusual test samples. The current study builds on this work, exploring the sources of overfitting and how they vary across linear regression contexts and presenting tools to address these limitations of PRESS.

Section \ref{Overfitting framework} formally characterizes two distinct mechanisms that cause regression models' performance to degrade when moving from training to test cases. First, the training observations' residuals appear in the least squares formula. Even if the true model lacks predictive power, this \emph{forbidden knowledge} produces an artificial correlation between the forecasts and actuals in the training sample. Second, a model has \emph{specialized training} to solve the types of forecasting problems it encountered in fitting. If a predictor $x_1$ varies widely in the training sample, with good numbers of both high- and low-valued observations, then the coefficient for $x_1$ will be precisely estimated. The fitted model is well-suited to explain the impacts of that wide variation in $x_1$ but will underperform on test samples that exhibit more variation in some other predictor $x_2$.

An estimator of out-of-sample error is introduced that is customized to address both sources of overfitting as they affect the particular test sample under consideration. Like PRESS, the derivations in this paper rely upon the popular ``leverage’’ variable that describes the extent to which each training observation influences the regression coefficients. An out-of-sample equivalent to this variable is introduced. Test cases with higher values for this new measure rely more heavily upon imprecisely estimated coefficients and are consequently more susceptible to overfitting.

The proposed estimator is evaluated in an extensive set of Monte Carlo simulations and also in an empirical application measuring the ability of Magnetic Resonance Imagery (MRI) brain scans to predict the results of a psychological test of behavior. Overfitting is observed in both contexts: predictions are less accurate out-of-sample than in-sample, particularly among models whose numbers of parameters are large relative to the size of the training sample. For settings in which the training and test samples are similar, both PRESS and the proposed estimator effectively correct for this bias.

The estimator proposed here performs more consistently than PRESS across different test samples and accurately forecasts model performance for samples that differ substantially from those encountered during fitting. Unlike PRESS, it requires data on the test cases' values for the explanatory variables. The most appropriate use case for the proposed approach is therefore at the time that the model is being deployed out-of-sample, when these predictor values are known.

The methods proposed in this study for projecting mean squared out-of-sample error as well as the out-of-sample equivalent to leverage have been made available in the R package \texttt{moose}, which can be downloaded from \citet{Rohlfs2022b}. The code used to produce the results in this paper is also provided with the supplementary materials. Additional notes about the derivations and specifications are also provided in an online appendix. A preprint version of this article can be found at \cite{Rohlfs2022a}.

\section{Conceptual Framework} \label{Overfitting framework}

There is a large population of $N$ units indexed by $i$, and there is some outcome variable $y_i$ that depends linearly upon a set of $k$ observable predictors $\bm{x_i} = [x_{1i}\: \cdots\: x_{ki}]$ and a set of unobservable factors represented collectively by a unit-specific residual $\epsilon_i$:

$$ y_i = \bm{x_i \beta} + \epsilon_i, $$

\noindent where the coefficient vector $\bm{\beta}$ consists of $k$ constant impacts $[\beta_{1}\: \cdots\: \beta_{k}]'$ of the $k$ predictors. The residuals are mean zero and distributed orthogonally to the predictors, so that $E[\epsilon_i | \bm{x_i}]=E[\epsilon_i]=0$ for all $i$. Individual units' residuals are distributed independently from one another but are not identically distributed.

A \emph{training sample} consisting of $n < N$ ``in-sample'' observations is drawn randomly from the population. Without loss of generality, suppose that these training observations constitute the first $n$ units in the population, so that the index $i$ simultaneously refers to the observations' respective positions in the population and in the training sample. The variables from this sample are represented by the vector of outcomes $\bm{Y} = [y_{1}\: \cdots\: y_{n}]'$, the $n$ by $k$ matrix of predictors $\bm{X} = [\bm{x_{1}'}\: \cdots\: \bm{x_{n}'}]'$, and the vector of true residuals $\bm{\epsilon} = [\epsilon_{1}\: \cdots\: \epsilon_{n}]'$. The vector $\bm{Y}$ and the matrix $\bm{X}$ of training data are used together to construct a vector $\bm{\hat{\beta}} = \bm{(X'X)^{-1}(X'Y)}$ of estimated coefficients via ordinary least squares (OLS). For any observation $i$ in the larger population, the forecasted outcome value $\bm{x_i \hat{\beta}}$ is denoted $\hat{y}_i$, and the estimated residual $y_i -\hat{y}_i$ is denoted $\hat{\epsilon}_i$.

Following the literature, let $\bm{H=X(X^\prime X)^{-1} X^\prime}$ denote the ``hat'' matrix, with elements $h_{ij}=\bm{x_{i} (X^\prime X)^{-1} x_{j}'},$ where the $i^{th}$ diagonal element $h_i$ represents the ``leverage'' or ``pull'' of an observation, indicating its influence on the estimated coefficients \citep{AngristPischke2009, HoaglinWelsch1978, Huber1975, MelounMilitky2010}. The hat matrix is symmetric and idempotent, and $|h_{ij}| \le 1$ for all $i,j$, $\sum_{j=1}^{n} h_{ij} = 1$ and $h_i \in [0,1]$ for all $i,$ and $\sum_{i=1}^{n} h_i = k$, and it follows from idempotence of $\bm{H}$ that each of these diagonal elements $h_i$ is equal to $\sum_{j=1}^{n} h_{ij}^2,$ the sum along row $i$ of the squared elements of the matrix.

In addition to the training sample, a \emph{test sample} consisting of $m$ ``out-of-sample'' observations is drawn randomly from the population and is used to evaluate the accuracy of the predictor. Again without loss of generality, these units are assigned indices $i$ from $n+1$ to $n + m$. The variables associated with the test sample are represented by the vector of outcomes $\bm{Y^o} = [y_{n+1}\: \cdots\: y_{n+m}]'$, the matrix of predictors $\bm{X^o} = [\bm{x_{n+1}'}\: \cdots\: \bm{x_{n+m}'}]'$, and the vector of true residuals $\bm{\epsilon^o} = [\epsilon_{n+1}\: \cdots\: \epsilon_{n+m}]'$. Forecasted values $\bm{\hat{Y}^o} = [\hat{y}_{n+1}\: \cdots\: \hat{y}_{n+m}]'$ for the test sample are computed as $\bm{X^o \hat{\beta}}$ using the parameter vector $\bm{\hat{\beta}}$ that is estimated from the training sample. A natural measure of forecasting accuracy is the out-of-sample mean squared error (MSE or $MSE^o$), which can be defined in the current context as $\frac{1}{m} \sum_{i=n+1}^{n+m} (y_i -\hat{y}_i)^2.$

A key aim in constructing a measure of $MSE^o$ is to use it as a criterion by which to select among competing models to forecast $y_i$ for out-of-sample cases $i>n.$ To that end, it is essential to consider the possibility that the model is incorrectly specified. In a standard formulation of the problem, misspecification arises if there is some omitted variable $z_i$ and some included predictor $x_{li}$ such that $\epsilon_i = \alpha*z_i + u_i$ and $z_{i} = \delta*x_{li} + \xi_i,$ where $u_i$ and $\xi_i$ are pure noise but $\alpha \ne 0$ and $\delta \ne 0.$ This characterization is highly general and, depending upon the type of omitted variable, can be used to express unobserved predictors, nonlinear functional forms, or omitted interactions with existing predictors. Supposing that the training and test samples are randomly drawn from the same population, the estimated coefficient $\hat{\beta}_l$ will exhibit the same bias of $\alpha * \zeta$ both in-sample and out-of-sample. Thus, this misspecification affects the interpretation of the estimated coefficient, but it does not impact the generalizability of the model's in-sample forecasts to out-of-sample cases. \citet{Shmueli2010} provides a detailed discussion of this distinction between explanation and prediction. When the training and test data are not randomly drawn from the same population, the task of generalizing results across samples is more complex and is beyond the scope of this study.

\subsection{Estimators of Out-of-Sample Error}

\subsubsection{PRESS} \label{Overfitting press}

A common measure of $MSE^o$ developed by \citet{Allen1971, Allen1974} is $1/n$ times the Predicted Residual Error Sum of Squares (PRESS), defined as the leave-one-out cross-validation error from the training data \citep{Myers2010, Tarpey2000}:

$$ PRESS = \sum_{i=1}^{n} (y_i -\hat{y}_{(i)})^2, $$

\noindent where $\hat{y}_{(i)}$ is the out-of-sample forecast of $y_i$ using all $n-1$ training observations except for $i.$ Typically, the computational complexity of leave-one-out cross-validation is prohibitively high, but for linear regression, \citet{Allen1974} demonstrates and \citet{SeberLee2003} show in a simplified proof that the estimated jackknife residual $\hat{\epsilon}_{(i)}$ equals $\hat{\epsilon}_i/(1-h_i).$ Thus, PRESS can be calculated solely from the full training sample regression without the computational cost of additional matrix inversions as:

\begin{equation} \label{press} PRESS = \sum_{i=1}^{n} (\frac{y_i -\hat{y}_i}{1-h_i})^2, \end{equation}

\noindent The ratio $\frac{PRESS}{n}$ measures the amount of error that will be observed on average when a linear model that was fitted on the $n$-observation training sample is deployed on new test cases. This statistic is only informative, however, for the average out-of-sample case. It does not identify which out-of-sample cases will have larger or smaller errors.

\subsubsection{Proposed Approach} \label{Overfitting proposed}

The remainder of this section introduces a new approach for projecting $MSE^o$. A formula is first derived for the non-stochastic case, when the test data have the same values for the predictors as the training data, and then more generally for the case in which out-of-sample cases have known but previously unseen predictor values.

\paragraph{Non-Stochastic Predictors} \label{Overfitting non-stochastic}

\mbox{}

The first source of overfitting bias considered here is \emph{forbidden knowledge}---the presence of the true residual $\epsilon_i$ in both the formula for the true outcome $y_i$ and the forecasted value $\hat{y}_i$, which causes forecasting errors to be generally smaller in magnitude for units in the training sample than for cases in the broader population. To isolate this mechanism, we consider a data generating process in which the predictors $\bm{x_i}$ are non-stochastic---as in the case of experimentally assigned dosages of medication in drug trials. In order to ensure comparability between the training and test samples, the sample sizes and matrices of predictors are set to be identical between the two datasets. Hence, $m = n$ and $\bm{X^o} = \bm{X}.$

In this non-stochastic predictor setting, the in-sample forecasted value $\hat{y}_i$ for each observation $i$ in the training sample is by construction identical to the out-of-sample forecasted value $\hat{y}_{j}$ for the corresponding observation $j=n+i$ in the test sample; \emph{i.e.,} $\bm{\hat{Y^o}} = \bm{\hat{Y}}$. By the formula for the OLS coefficients, this vector can be expressed in terms of the hat matrix as $\bm{HY}$. The vectors $\bm{Y}$ and $\bm{Y^o}$ of actual outcomes differ due to distinct vectors of true residuals $\bm{\epsilon}$ and $\bm{\epsilon^o}$. Let $\bm{\Omega}$ and $\bm{\Omega^o}$ denote the variance-covariance matrices $E[\bm{\epsilon\epsilon'}]$ and $E[\bm{\epsilon^o\epsilon^{o\prime}}]$ of these vectors, which are known to be diagonal matrices as per the assumption of mutually independent errors. Let the $i^{th}$ diagonal elements of these matrices be denoted $\sigma^2_i$. Any two cases with identical $\bm{x_i}$ vectors are observationally equivalent, and their distributions of possible residual values are indistinguishable by the researcher. Thus, without loss of generality and in keeping with the literature \citep{White1980}, suppose that such cases have identical residual variance. It follows that $\sigma_{n+i}^2 = \sigma_i^2$ for all $i = 1,...,n$ and consequently that $\bm{\Omega^o=\Omega}$.

Substituting $\bm{X\beta+\epsilon}$ for $\bm{Y}$, the vector $\bm{\hat{Y}}=\bm{HY}$ of in-sample forecasts can be expressed as $\bm{X(X'X)^{-1}X'X\beta + X(X'X)^{-1}X'\epsilon}$, which simplifies to $\bm{X\beta + H\epsilon}$. The vector $\bm{\hat{\epsilon}} = \bm{Y-\hat{Y}}$ of in-sample forecasting errors can then be re-written as $\bm{X\beta + \epsilon -X\beta -H\epsilon}=\bm{(I_n -H)\epsilon}$. The corresponding vector $\bm{\hat{\epsilon}^o} = \bm{Y^o-\hat{Y}}$ of out-of-sample forecasting errors simplifies to $\bm{\epsilon^o-H\epsilon}$. The expected sums of squares for these estimated residuals can be expressed as follows:

\begin{equation} \label{inSampleV} \begin{split} E[\bm{(Y-\hat{Y})^\prime(Y-\hat{Y})}] = E[\bm{\epsilon'(I_n-H)\epsilon}] = \sum_{i=1}^n (1-h_i)*\sigma_i^2, \:\textrm{and} \end{split} \end{equation}

\begin{equation} \label{outSampleV} \begin{split} E[\bm{(Y^o-\hat{Y})^\prime(Y^o-\hat{Y})}] = E[\bm{\epsilon^{o\prime}\epsilon^o +\epsilon'H\epsilon}] = \sum_{i=1}^n (1+h_i)*\sigma_i^2. \end{split} \end{equation}

\noindent Equation \ref{inSampleV} simplifies $\bm{(I_n-H)'(I_n-H)}$ to $\bm{I_n-H}$ by symmetry and idempotence of the residualization matrix $\bm{I_n-H}$, and Equation \ref{outSampleV} employs the same simplification for $\bm{H}$. Both equations simplify $E[\bm{\epsilon'H\epsilon}]$ to $E[\sum_{i=1}^n h_i \epsilon_i^2],$ which follows from the assumption that $E[\epsilon_i\epsilon_j]=0$ for all $i \ne j$, a property that also causes $E[\bm{\epsilon^{o\prime}H\epsilon}]$ and $E[\bm{(H\epsilon)'\epsilon^o}]$ to drop out in Equation \ref{outSampleV}. The last step in Equation \ref{outSampleV} also employs $\sigma_{n+i}^2 = \sigma_i^2$ for all $i.$

If the true coefficient vector $\bm{\beta}$ were known, the expected in-sample and out-of-sample sums of squared residuals would both equal $\sum_{i=1}^n \sigma_i^2.$ When the training data are used, the expected sum is lower than that amount by the proportion $h_i \in [0,1]$ for each observation $i$ due to the presence of the vector $\bm{\epsilon_I}$ of in-sample residuals in the formulas for $\bm{\hat{\beta}}$ and consequently $\bm{\hat{Y}}$---the ``forbidden knowledge'' source of overfitting. The resulting expected squared estimated residual for out-of-sample case $n+i$ exceeds $\sigma_i^2$ by the proportion $h_i$ for each $i$ due to the additional independent error associated with the estimated coefficients being used. Hence, the forecasting error will tend to have lower variance in-sample than out-of-sample.

The product of $h_i*\sigma^2_i$ in both Equations \ref{inSampleV} and \ref{outSampleV} indicates that the magnitude of the overfitting bias is largest when the estimates with the greatest leverage---in the sense of having exceptional values for some predictors---are also cases for which the residuals have high variance. Because the average value of $h_i$ rises with the number of regressors $k$, the magnitude of the bias also tends to increase with model complexity. Holding the number of parameters constant, $\lim_{n\to\infty} \frac{1}{n}\sum_{i=1}^n h_i\sigma^2_i=0$; hence, for this case of a matrix $\bm{X}$ of non-stochastic predictor values that is the same between the training and test datasets, the problem of overfitting is not applicable when the sample size is large relative to the number of parameters being estimated. Even for samples that are moderate-sized relative to the parameters, however, the bias could be large for observations with extreme values for the predictors and consequently high values for $h_i$.

As per the equations above, the square of each in-sample estimated residual can be divided by $1-h_i$ to produce an unbiased estimator of $\sigma^2_i$. This estimator of $\sigma^2_i$, originally proposed by \citet{MacKinnonWhite1985} is a modified version of the original unscaled and downward-biased variant by \citet{White1980}. Alternative estimators of $\sigma^2_i$ vary in their small sample performance in different contexts and the computational intensity of the calculations \citep{Cribari-Neto2004, DavidsonMacKinnon1993, HausmanPalmer2012, MacKinnonWhite1985}. In tests not shown here, a few variants were considered and found to be generally comparable. The factor of $1/(1-h_i)$ is used because it is relatively simple and aligns closely with the derivation above. Substituting from Equations \ref{inSampleV} and \ref{outSampleV} produces the following estimator for $MSE^o$:

\begin{equation} \label{MSEest} \hat{MSE}^o_{Non{\text -}Stochastic} = \frac{1}{n} \sum_{i=1}^n (\frac{1+h_i}{1-h_i})\hat{\epsilon}_i^2 \end{equation}

\noindent By $h_i \in [0,1]$ for all $i$, the adjusted squared errors all equal or exceed the unadjusted ones, with the largest upward scaling occurring for high leverage observations.

\paragraph{Stochastic Predictors} \label{Overfitting stochastic}

\mbox{}

In most real world applications, the values of the predictors are outside of the researcher's control, and $\bm{X}$ and $\bm{X^o}$ do not match. The predicted values for the test observations can be described as follows:

\begin{equation} \label{Yohat} \bm{\hat{Y}^o} = \bm{X^o(X'X)^{-1}X'Y} \end{equation}

\noindent Let $\bm{H^o}$ denote the out-of-sample hat matrix $\bm{X^o(X'X)^{-1}X'}$, which pre-multiplies the in-sample outcome vector $\bm{Y}$, as on the right-hand side of Equation \ref{Yohat}, resulting in out-of-sample projections on the left-hand side, and its structure highlights two distinct aspects of an observation's ``leverage.'' If in-sample unit $i \le n$ has an especially high or low value for some predictor $x_{1i}$ in the in-sample matrix $\bm{X}$, then $i$'s residual $\epsilon_i$ will be particularly influential in the determination of $\hat{\beta}_1$, the estimated coefficient for that predictor. In the out-of-sample matrix $\bm{X^o}$, if the value $x_{1j}$ is especially high or low for that same predictor for some unit $j > n$, then unit $j$'s forecasted value $\hat{y}_j$ will rely heavily upon the estimated coefficient $\hat{\beta}_1$, and imprecision in this estimated coefficient will have an outsized contribution to the forecast $\hat{y}_j$.

For each test case $j>n$, the matrix $\bm{H^o}$ has a row $\bm{h^{o}_j} = [h^o_{j1}\: \cdots\:h^o_{jn}]$, and the out-of-sample forecast $\hat{y}_j = \sum_{i=1}^{n} h^o_{ji}y_i$ is a weighted sum of the in-sample actuals. Like in training, the weights sum to one, but unlike in training, the magnitudes of the weights $h^o_{ji}$ can exceed one. $\bm{H^o}$ is not idempotent, symmetric, or square, so $\bm{H^{o\prime}H^o}$ does not simplify, and there is not a diagonal element that can serve as the out-of-sample equivalent to the leverage variable $h_i.$ Nevertheless, the sum $\sum_{i=1}^{n} h_{ji}^{o^2}$ of squared elements of row $j$ represents a measure of concentration of the weights, indicating the extent to which the prediction $\hat{y}_j$ relies upon the values from a small number of training cases. As noted earlier in this section, for an in-sample case $i,$ this sum of squares is identical to $h_i$; it is thus well-suited to serve as the out-of-sample equivalent to leverage and will henceforth be denoted $h_i$ for all $i$.

The second source of overfitting described in Section \ref{Overfitting introduction}, \emph{specialized training}, arises when some explanatory variables exhibit unusually little variation due to the idiosyncracies of the training sample, and their coefficients are imprecisely estimated. This deficiency will tend not to matter for in-sample performance. The imprecise coefficient estimates are not used extensively for in-sample projection, because the training sample has few movements in those predictors that require explaining. The test sample is unlikely to have the same idiosyncracies, however, and it may make extensive use of those coefficients, leading to imprecision in out-of-sample forecasting. This source of overfitting is most severe when the number of training cases is low relative to the number of estimated parameters.

At the time the model is deployed out-of-sample, the matrix $\bm{X^o}$ of out-of-sample predictor values is known, but the out-of-sample outcome vector $\bm{Y^o}$ is not. The variance-covariance matrix of the forecasting error is slightly modified from the expression for non-stochastic predictors in Equation \ref{outSampleV}:

\begin{equation} \label{stochastic_outSampleV} \begin{split} E[\bm{(Y^o-\hat{Y})^\prime(Y^o-\hat{Y})}] = E[\bm{\epsilon^{o\prime}\epsilon^o +\epsilon'H^{o\prime}H^o\epsilon}] = \sum_{j=n+1}^{n+m} \sigma_j^2 + \sum_{j=1}^{m} \sum_{i=1}^{n} (h^{o}_{ji} \sigma_i)^2. \end{split} \end{equation}

\noindent As in Equation \ref{outSampleV}, the first term in Equation \ref{stochastic_outSampleV} describes the error associated with the out-of-sample residuals, and the second term results from imprecision in training---with in-sample error variances weighted by their importance in the out-of-sample projection.

With $\bm{X}$ and $\bm{X^o}$ different, $\sigma_i^2$ is no longer an appropriate approximation for $\sigma_{n+i}^2$. Equation \ref{auxiliary} below specifies a variant of the auxiliary regression from \citet{White1980} for $\sigma_i^2$ that depends upon the same set of predictors as in the main regression equation:

\begin{equation} \label{auxiliary} \sigma^2_i = \bm{x_i \gamma} + \nu_i, \end{equation}

\noindent where $\nu_i$ is a mean zero residual. As in Subsection \ref{Overfitting non-stochastic} above, for in-sample cases $i \le n$ observed in the training data, $\hat{\epsilon}^{2}_i/(1-h_i)$ can be used as an estimable proxy for $\sigma^2_i$. This substitution introduces measurement error into the dependent variable of the auxiliary regression in Equation \ref{auxiliary} but does not bias the coefficient estimates. Let $\bm{\hat{\epsilon}^{*2}} = [\hat{\epsilon}^{*2}_{1}\: \cdots\: \hat{\epsilon}^{*2}_{n}]'$ denote the vector of squared scaled estimated residuals for the training sample. The OLS estimator $\bm{\hat{\gamma}}$ for the auxiliary coefficient vector is then expressed as follows:

$$ \bm{\hat{\gamma}} = \bm{(X'X)^{-1}X'\hat{\epsilon}^{*2}}. $$

\noindent The $m$ out-of-sample residual variances are forecasted by the vector $\bm{X^o\hat{\gamma}=H^o\hat{\epsilon}^{*2}}$. Thus, the sum $\sum_{j=n+1}^{n+m} \sigma_j^2$ from Equation \ref{stochastic_outSampleV} can be estimated as $\sum_{j=1}^{m} \sum_{i=1}^{n} h^o_{ji} \hat{\epsilon}^{*2}_I$. The second term in that expression can be estimated by substituting $\hat{\epsilon}^{*2}_i$ for $\sigma_i^2$ for each $i$. The out-of-sample MSE can then be projected as:

\begin{equation} \label{stochastic_MSEest} \hat{MSE}^o_{Stochastic} = \frac{1}{m} \sum_{j=1}^{m}\sum_{i=1}^{n}(h^o_{ji}+h_{ji}^{o^2})\hat{\epsilon}^{*2}_i \end{equation}

\noindent As the double summations in Equation \ref{stochastic_MSEest} show, each training observation carries some potentially non-zero weight in the forecasted squared error for each of the test cases being forecasted. For the special case in which $\bm{X^o = X}$, the aforementioned properties of $\bm{H}$ apply so that Equation \ref{stochastic_MSEest} is identical to the non-stochastic variant in Equation \ref{MSEest}.

\section{Evaluation Framework} \label{Overfitting evaluation}

\subsection{Simulation Design} \label{Overfitting simulation}

Simulated data are generated in which, in the baseline configuration, the true model has 45 predictors with standard deviations of 45, 44,..., down to one, all with the same true effect of $\beta$ on the outcome variable $y_i$; the true intercept is zero. The residual $\epsilon_i$ has standard deviation of $150*\beta$, and $\beta$ is set so that the variance of $y_i$ is one. The predictors are all independent normals. The residual is a product of normals whose variance increases with each of the predictors. In each of 1,000 simulation iterations, an 80-observation training sample is generated. Using that training sample, 45 OLS regressions are estimated that include the $k$ highest variance predictors, where $k=1,...,45$; each regression includes an intercept. Also in each iteration, a 10,000-observation test sample is generated with fresh draws of each of the predictors and residuals. Each of the 45 models is used to produce forecasted values $\hat{y}_j$ for $j>n$ so that 45 model-specific $MSE^o$ values are estimated. For each iteration and model, both $PRESS/n$ and the proposed measure of out-of-sample error are compared against these actual values of $MSE^o$.

\begin{table}[ht!]
\renewcommand{\arraystretch}{2}
\centering
\resizebox{0.75\textwidth}{!}{
\begin{tabular}{c|c|c|c|c|c|c|c}
\multicolumn{8}{c}{Panel A: Monte Carlo simulation} \\
\shortstack{Predictors\\ in\\ True\\ Model} & \shortstack{Predictors\\ in\\ Estimated\\ Model} & \shortstack{Error\\ Process} & \shortstack{Training\\ Sample\\ Size} & \shortstack{Test\\ Sample\\ Size} & \shortstack{Out-of-sample\\ Predictor\\ Design} & Iterations & \shortstack{Total\\ Regressions} \\
\hline
\rule{0pt}{8ex} 45 & \shortstack{1, ..., 45} & \shortstack{Homoskedastic\\ or\\ Heteroskedastic} & \shortstack{50, 80, 100, 125,\\ 250, 500, and 1,000} & 10,000 & \shortstack{Non-stochastic\\ or\\ Stochastic} & 1,000 & 630,000 \\
\hline
\rule{0pt}{8ex} 950 & \shortstack{1, 50,\\ 100, ..., 950} & \shortstack{Homoskedastic\\ or\\ Heteroskedastic} & \shortstack{1,000, 1,250, 1,500,\\ 1,750, 2,000, 2,500,\\ and 5,000} & 10,000 & \shortstack{Non-stochastic\\ or\\ Stochastic} & 1,000 & 280,000 \\
\end{tabular}}
\resizebox{0.50\textwidth}{!}{
\begin{tabular}{c|c|c|c|c}
\multicolumn{5}{c}{Panel B: Neurocognitive Aging Data} \\
\shortstack{Predictors\\ in\\ Estimated\\ Model} & \shortstack{Test\\ Sample\\ Size} & \shortstack{Test\\ Sample\\ Size} & Iterations & \shortstack{Total\\ Regressions} \\
\hline
\rule{0pt}{8ex} 1, ..., 50 & \shortstack{25\% ($\sim 71$),\\ 50\% ($\sim 142$),\\ and 75\% ($\sim 212$)} & \shortstack{75\% ($\sim 212$),\\ 50\% ($\sim 142$),\\ and 25\% ($\sim 71$)} & 1,000 & 150,000 \\
\end{tabular}}
\caption{Alternate Configurations for Monte Carlo Simulation and Empirical Application}
    \label{tab:configurations}
\end{table}

In addition to this baseline configuration with 45 true predictors, heteroskedastic errors, an 80-observation training sample, and stochastic predictors, a variety of alternate simulation specifications are run with different numbers of predictors, training sample sizes, and error processes as described in panel A of Table \ref{tab:configurations}. These alternate specifications help to illustrate the conditions under which overfitting bias is likely to arise as well as the accuracy of the proposed estimator of $MSE^o$ in these different contexts.

As a preview of the findings across these configurations, the proposed method and $PRESS/n$ are both most accurate when overfitting is low due to large training samples, small numbers of estimated parameters, or, for the proposed method, when there is no specialized training component because the predictors are non-stochastic. Among the complex models with small training samples, both estimators produce less accurate measures of $MSE^o$; they continue to perform comparably on average, but the proposed approach shows greater consistency across iterations and greater visibility into case-specific variation.

\subsection{Neurocognitive Aging Data} \label{Overfitting application}

In the empirical portion of this study, regressions are estimated to predict the score from a psychological test of executive function, a measure of memory, mental flexibility, and self-regulation, as a function of the physical structure of subjects' brains. Psychological tests and MRI scans of 283 subjects from Ithaca and Toronto are taken from the Neurocognitive Aging Data Release \citep{Spreng2022}. This choice of this application, with a cost of \$3,000 per subject for the full dataset and many more potential predictors than observations, highlights the importance in some situations of effectively using every training case and the difficulty of determining the appropriate level of model complexity.

The executive index score has mean zero and variance 0.47, with values ranging from -2.03 to +1.80. Downsampling is applied to anatomical MRI images to produce mean grayscale values for $32*32*16 = 16,384$ voxels or cells, whose top 50 principal components are kept as candidate predictors.

In each of 1,000 iterations, cases from the 283-observation dataset are randomly selected with 50\% probability into the training set. Using that training sample, 50 OLS regressions are estimated that include the $k=1,...,50$ top eigenvectors from the MRI data; each regression includes an intercept. For each iteration and estimated model, out-of-sample projections $y_j$ for $j>n$ are generated for the remaining observations to estimate $MSE^o$. $PRESS/n$ and the proposed estimator are constructed and compared with the true out-of-sample MSE for each iteration and model. The process is also performed for 25\% and 75\% training sets, as described in panel B of Table \ref{tab:configurations}. 

\section{Results} \label{Overfitting results}

\subsection{Average Test Sample Performance} \label{Overfitting full results}

\begin{figure}[ht!]
\begin{subfigure}{.5\textwidth}
  \centering
  \includegraphics[width=.8\textwidth]{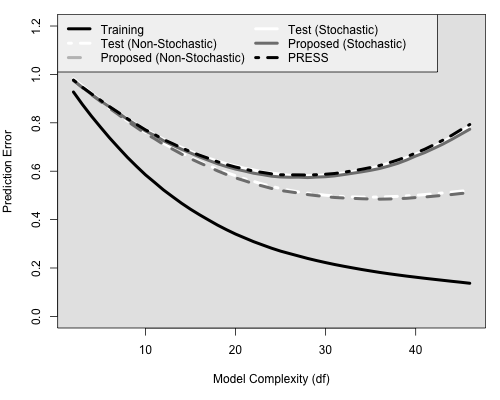}
  \caption{Monte Carlo simulation}
  \label{fig:rss-simulation}
\end{subfigure}
\begin{subfigure}{.5\textwidth}
  \centering
  \includegraphics[width=.8\textwidth]{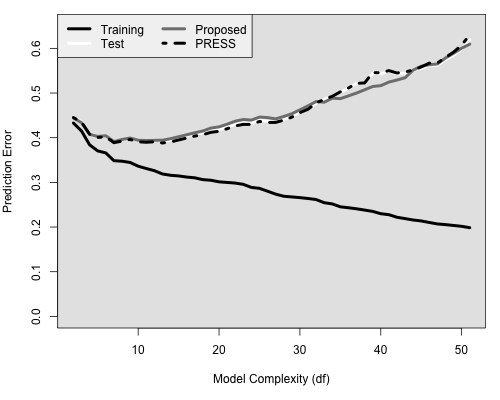}
  \caption{Neurocognitive Aging Data}
  \label{fig:rss-empirical}
\end{subfigure}
\caption{Mean training, test, and projected test errors by model complexity}
\label{fig:rss}
\end{figure}

The main results are presented for the baseline specifications of the simulation and empirical application in Figure \ref{fig:rss}, whose layout mirrors that of Figure 7.1 from \citet{HastieTibshiraniFriedman2009}. Each line shows an MSE as it varies with model complexity (the number of included predictors plus one for the intercept). In Figure \ref{fig:rss-empirical}, both $PRESS/n$ and the proposed measure are artificially close to the test MSE due to the sampling design whereby each observation appears alternately in the training and test data in different iterations. As the results in the graphs show, prediction error for the training samples (shown in black) declines steadily as the number of predictors rises. In the test samples, the variance of the estimated residuals (shown in white) is consistently higher than in the training sample and follows a U-shape, initially declining and then rising with model complexity, so that training and test error rates diverge as model complexity rises. Both the proposed measure and $PRESS/n$ track test error closely.

\begin{figure}[ht]
\begin{subfigure}{.5\textwidth}
  \centering
  \includegraphics[width=.8\textwidth]{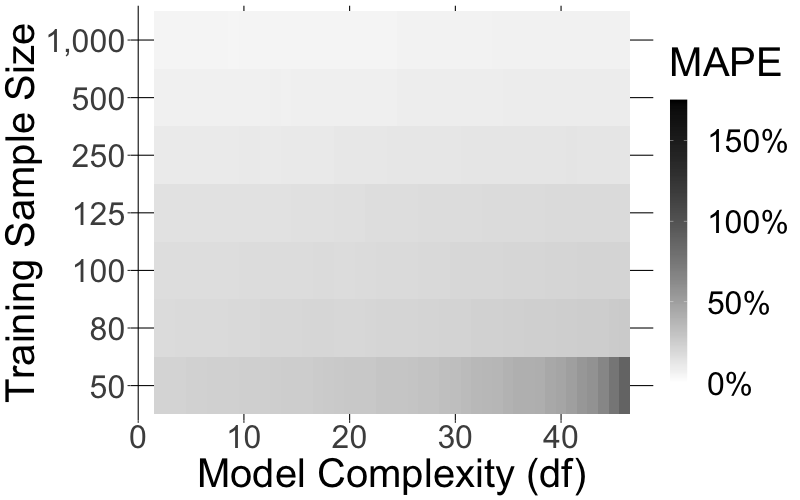}
  \caption{Proposed approach, simulations}
  \label{fig:mape-proposed-simulation}
\end{subfigure}
\begin{subfigure}{.5\textwidth}
  \centering
  \includegraphics[width=.8\textwidth]{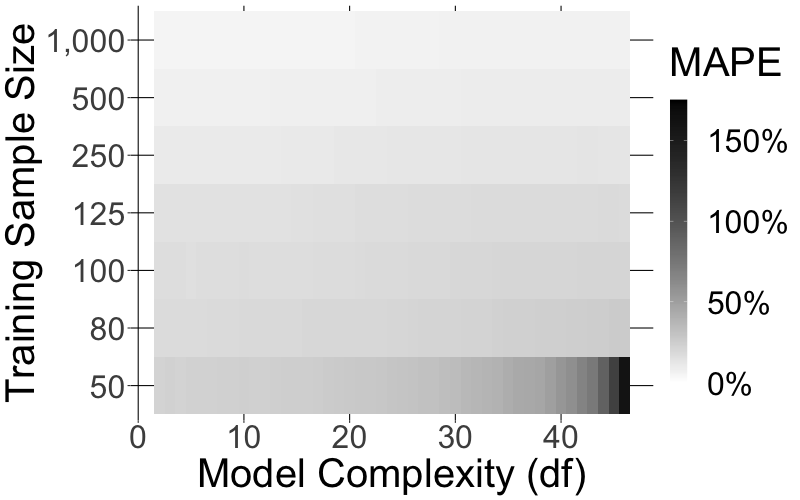}
  \caption{$PRESS/n$, simulations}
  \label{fig:mape-press-simulation}
\end{subfigure}
\begin{subfigure}{.5\textwidth}
  \centering
  \includegraphics[width=.8\textwidth]{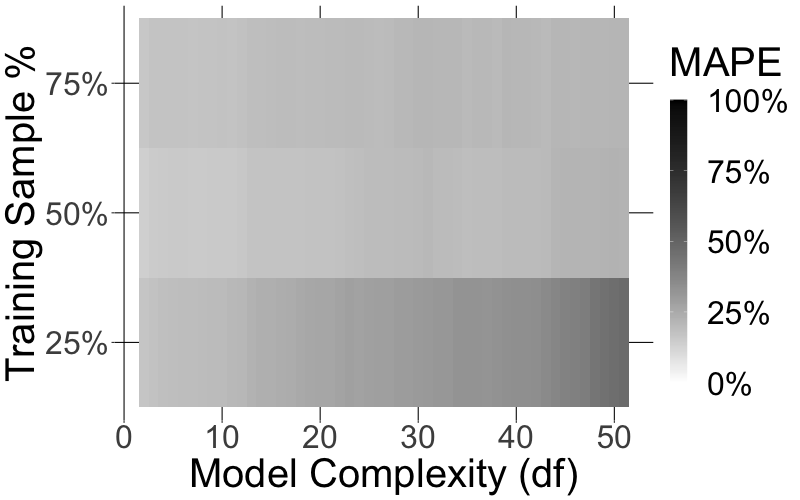}
  \caption{Proposed, Neurocognitive Aging Data}
  \label{fig:mape-proposed-empirical}
\end{subfigure}
\begin{subfigure}{.5\textwidth}
  \centering
  \includegraphics[width=.8\textwidth]{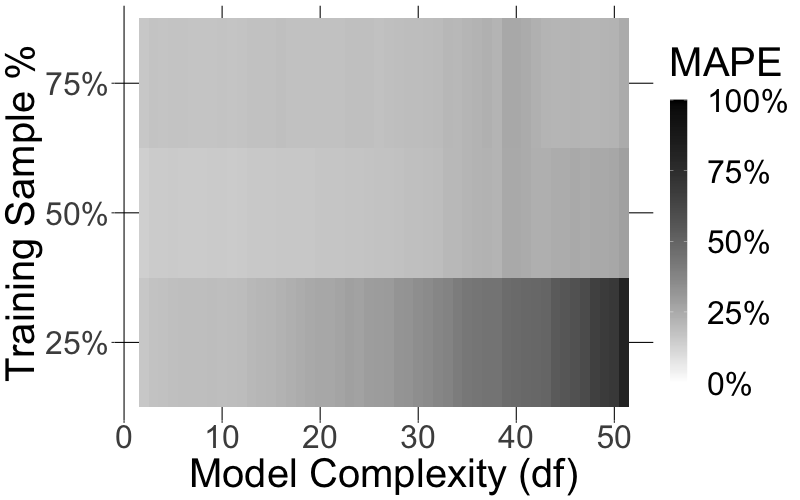}
  \caption{$PRESS/n$, Neurocognitive Aging Data}
  \label{fig:mape-press-empirical}
\end{subfigure}
\caption{MAPE of out-of-sample mean squared error forecasts for proposed measure and $PRESS/n$}
\label{fig:mape}
\end{figure}

The level of shading of each cell in Figure \ref{fig:mape} indicates a Mean Absolute Percentage Error (MAPE) between projected and actual estimated $MSE^o$. Means are taken across the 1,000 iterations of the simulation in the top panels and of the training/test splits in the bottom panels and are shown separately by training sample size, model complexity, proposed versus $PRESS/n$, and simulation versus empirical. As the results from Figure \ref{fig:mape} show, across iterations, MAPEs between actual and projected $MSE^o$ decrease with training sample size and increase with model complexity---so that the shading is lightest in the top left and darkest in the bottom right of each graph.

For the most overfit models, MAPEs are lower for the proposed measure than for $PRESS/n$; for the least overfit models, the MAPEs are similar between the two measures. The average MAPEs for the proposed out-of-sample MSE measure are 16.6\% across the simulation configurations, with values ranging from 5.1\% in the top left to 87.8\% in the bottom right. For $PRESS/n$ in simulations, the average MAPE is 17.2\%, and the values range from 5.2\% to 159.8\%. Across the specifications of the empirical application, the MAPEs for the proposed out-of-sample MSE average 22.4\%, ranging from 13.6\% to 47.1\%, and the MAPEs for $PRESS/n$ average 24.7\%, ranging from 13.6\% to 81.6\%.

\subsection{Variation Across Test Cases} \label{Overfitting variation}

\begin{figure}[ht]
\begin{subfigure}{.5\textwidth}
  \centering
  \includegraphics[width=.8\textwidth]{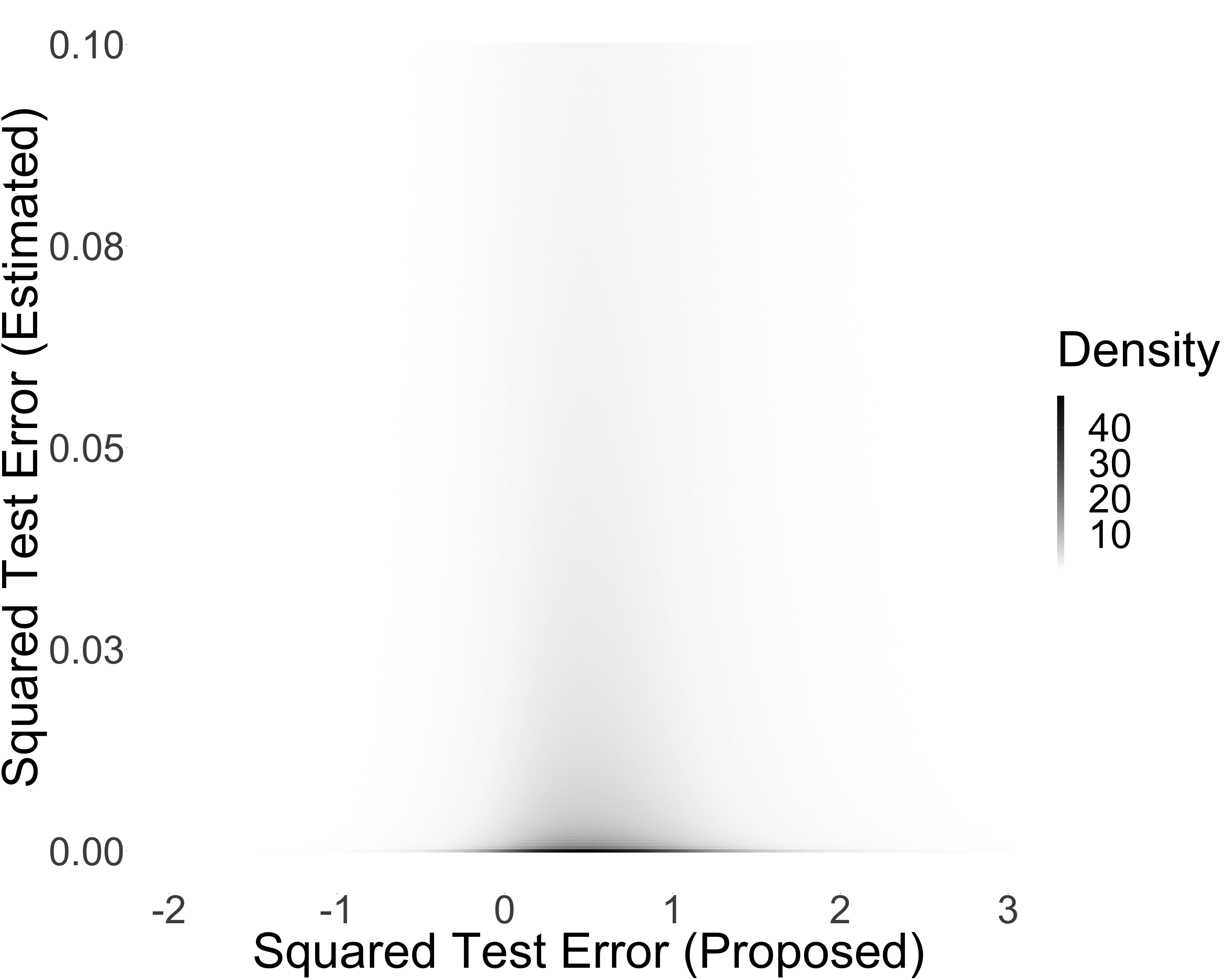}
  \caption{Monte Carlo simulation}
  \label{fig:joint-simulation}
\end{subfigure}
\begin{subfigure}{.5\textwidth}
  \centering
  \includegraphics[width=.8\textwidth]{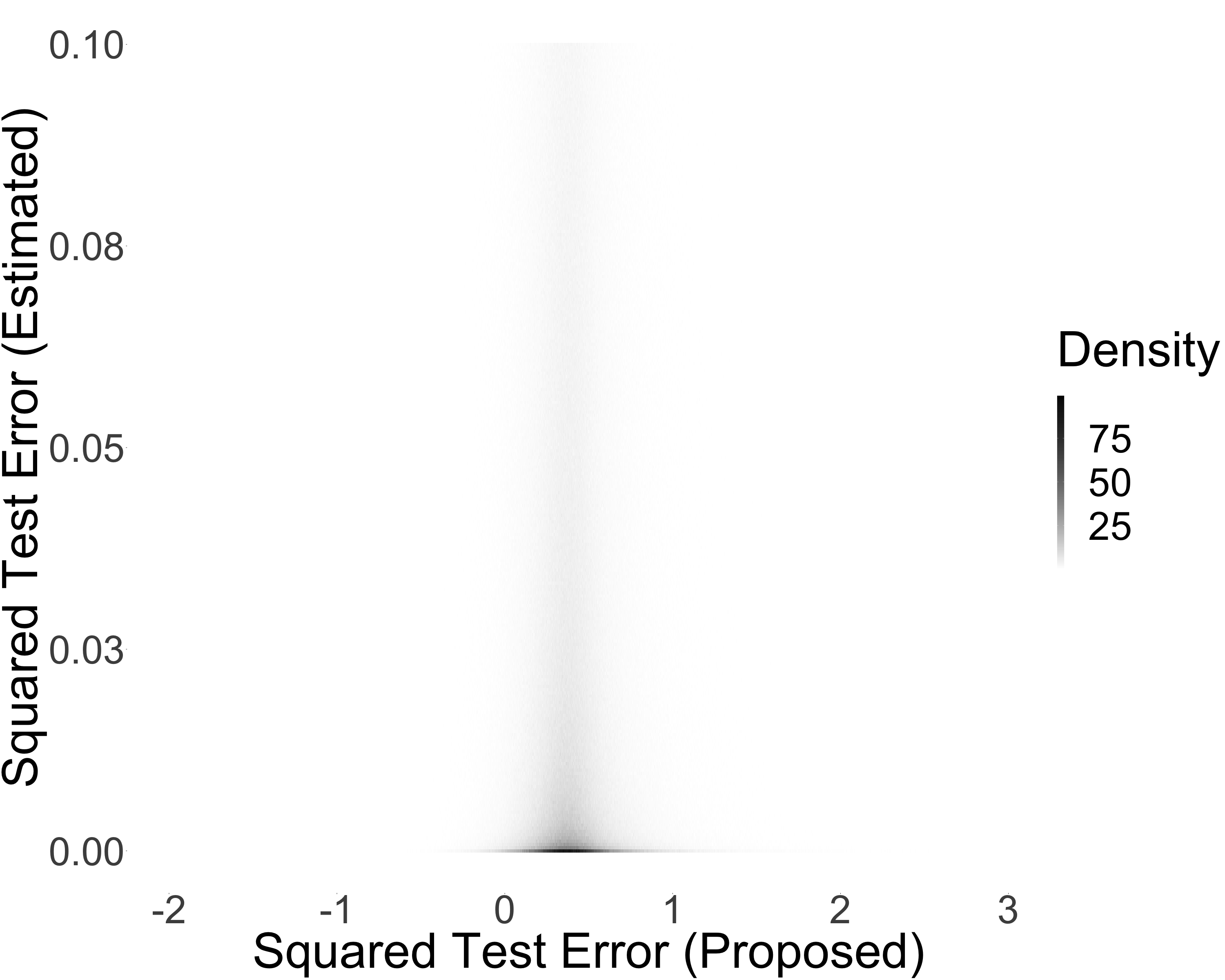}
  \caption{Neurocognitive Aging Data}
  \label{fig:joint-empirical}
\end{subfigure}
\caption{Joint densities of estimated (actual) and projected squared out-of-sample errors}
\label{fig:joint}
\end{figure}

Unlike PRESS, the proposed measure of out-of-sample error produces values for specific out-of-sample observations. To better understand the accuracy of these case-specific projections, Figure \ref{fig:joint} presents joint densities of the projected and actual out-of-sample squared errors across iterations, test cases, and levels of model complexity.

This variation across observations follows discernible patterns but exhibits some unintuitive results such as negative values. The standard deviations in Figure \ref{fig:joint} are larger for actuals than for projections: 1.36 versus 0.60 for the simulation and 0.89 versus 0.47 for the empirical application, because the actuals incorporate variation from the out-of-sample residuals, while the projections only reflect the variation generated by the predictors and the estimated coefficients. The actuals are positively skewed with most density concentrated around zero. Equation \ref{stochastic_MSEest} allows for negative projected squared residuals, which occurs in 8.6\% of simulated and 4.2\% of empirical cases.

\begin{figure}[ht]
\begin{subfigure}{.5\textwidth}
  \centering
  \includegraphics[width=.8\textwidth]{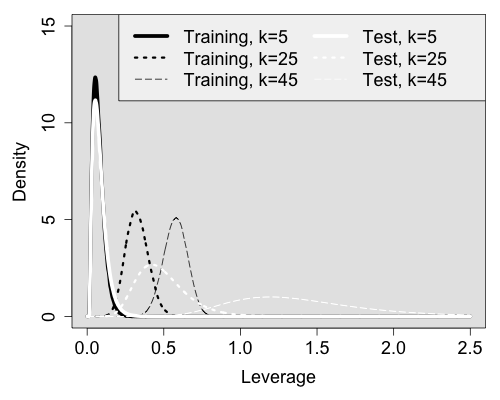}
  \caption{Monte Carlo simulation}
  \label{fig:hat-simulation}
\end{subfigure}
\begin{subfigure}{.5\textwidth}
  \centering
  \includegraphics[width=.8\textwidth]{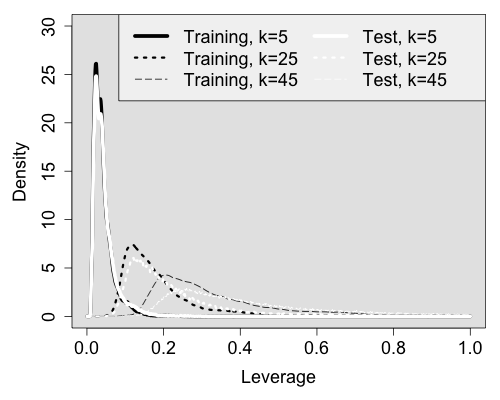}
  \caption{Neurocognitive Aging Data}
  \label{fig:hat-empirical}
\end{subfigure}
\caption{Density of leverage values in training and test samples by model complexity}
\label{fig:hat}
\end{figure}

The out-of-sample leverage variable defined in subsection \ref{Overfitting stochastic} provides a compact way to identify the extent to which different test cases deviate from the training sample in their values for the predictors. Aggregating by leverage $h_i$ makes it possible to learn from the variation across cases while avoiding the strange results that arise at the highest levels of granularity. Figure \ref{fig:hat} presents kernel-smoothed densities of the leverage variable, which is defined as the sum of squared elements of the hat matrix $\bm{H}$ or $\bm{H^o}$ for that observation. Densities are plotted for both the training and test samples for three different levels of model complexity, where the densities are taken across iterations and observations. The parameter $k$ in the legend indicates the number of predictors in the estimated model. As the graphs show, the mean and variance of leverage values both rise with model complexity and are higher for test than for training samples. Additionally, while $h_i$ cannot exceed one for training cases, it does exceed one for a non-trivial portion of test cases.

\begin{figure}[ht]
\begin{subfigure}{.5\textwidth}
  \centering
  \includegraphics[width=.8\textwidth]{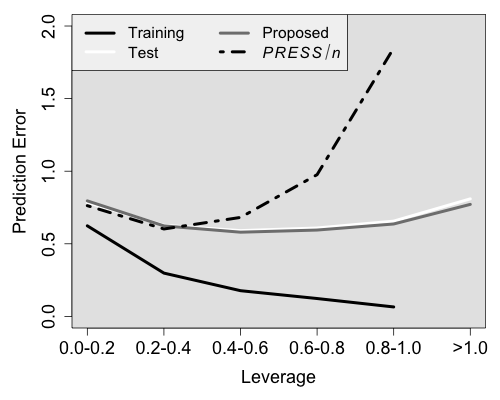}
  \caption{Monte Carlo simulation}
  \label{fig:hat-error-simulation}
\end{subfigure}
\begin{subfigure}{.5\textwidth}
  \centering
  \includegraphics[width=.8\textwidth]{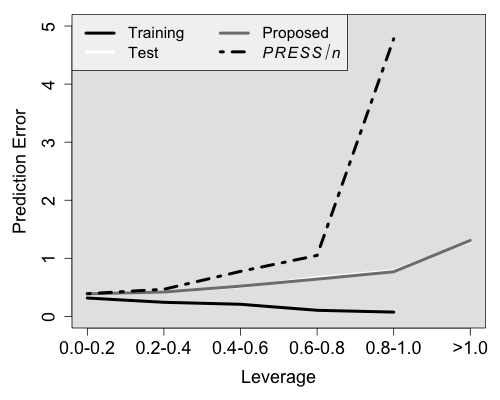}
  \caption{Neurocognitive Aging Data}
  \label{fig:hat-error-empirical}
\end{subfigure}
\caption{Actual and forecasted prediction error versus leverage}
\label{fig:hat-error}
\end{figure}

Figure \ref{fig:hat-error} shows how the different measures of out-of-sample prediction error perform for observations with different amounts of leverage, with $k$ ranging from 1 to 45 in Figure \ref{fig:hat-error-simulation} and from 1 to 50 in Figure \ref{fig:hat-error-empirical}. The proposed measure is computed for different values of leverage in the test samples, and $PRESS/n$ is computed for different values of leverage in the training samples. By $h_i \in [0,1]$, no training cases are observed with leverage greater than one. The results show that the gap between training and test error is consistently positive and widens as leverage rises. The proposed measure of $MSE^o$ consistently tracks the actual test error---and in both graphs, the gray and white lines are right on top of each other. By contrast, $PRESS/n$ does not provide a reliable means of forecasting out-of-sample MSE for these different subsamples. In addition to being undefined for the $>1.0$ category, $PRESS/n$ deviates considerably from the actual out-of-sample MSE at moderate and high leverage levels.

\begin{table}[ht!]
\renewcommand{\arraystretch}{1}
\begin{threeparttable}
\centering
\resizebox{0.75\textwidth}{!}{
\begin{tabular}{ccc|cccc|ccc}
\multicolumn{10}{c}{Panel A: Monte Carlo simulation} \\
\multicolumn{3}{c}{} & \multicolumn{4}{c}{All Cases} & \multicolumn{3}{c}{Leverage > 1.0} \\
\shortstack{Error\\ Type} & \shortstack{Predict-\\ ors in\\ True\\ Model} & \shortstack{Size of\\ Train-\\ ing\\ Sample} & \shortstack{Esti-\\ mated\\ Train-\\ ing\\ Error} & \shortstack{Esti-\\ mated\\ Test\\ Error} & $PRESS/n$ & \shortstack{Pro-\\ posed} & \shortstack{\% of Test\\ Cases} & \shortstack{Esti-\\ mated\\ Test\\ Error} & \shortstack{Pro-\\ posed} \\
\hline
\multirow{14}{*}{\shortstack{Homo-\\ sked-\\ astic}} & \multirow{7}{*}{45} & 50 & 0.326 & 1.511 & 1.684 & 1.501 & 46.64\% & 2.218 & 2.202 \\
 & & 80 & 0.414 & 0.802 & 0.804 & 0.800 & 14.62\% & 0.992 & 0.986 \\
 & & 100 & 0.447 & 0.728 & 0.735 & 0.732 & 2.71\% & 0.911 & 0.909 \\
 & & 125 & 0.467 & 0.681 & 0.683 & 0.682 & 0.11\% & 0.883 & 0.879 \\
 & & 250 & 0.512 & 0.611 & 0.610 & 0.610 & 0.00\% & & \\
 & & 500 & 0.534 & 0.583 & 0.581 & 0.582 & 0.00\% & & \\
 & & 1,000 & 0.547 & 0.570 & 0.570 & 0.570 & 0.00\% & & \\
 \cline{2-10}
 & \multirow{7}{*}{950} & 1,000 & 0.252 & 0.706 & 0.711 & 0.707 & 47.55\% & 0.747 & 0.750 \\
 & & 1,250 & 0.271 & 0.493 & 0.494 & 0.493 & 35.05\% & 0.348 & 0.348 \\
 & & 1,500 & 0.283 & 0.448 & 0.447 & 0.447 & 22.55\% & 0.271 & 0.271 \\
 & & 1,750 & 0.292 & 0.425 & 0.424 & 0.424 & 10.05\% & 0.239 & 0.239 \\
 & & 2,000 & 0.299 & 0.410 & 0.410 & 0.410 & 0.30\% & 0.226 & 0.227 \\
 & & 2,500 & 0.308 & 0.393 & 0.393 & 0.393 & 0.00\% & & \\
 & & 5,000 & 0.327 & 0.366 & 0.366 & 0.366 & 0.00\% & & \\
\hline
\multirow{14}{*}{\shortstack{Hetero-\\ sked-\\ astic}} & \multirow{7}{*}{45} & 50 & 0.291 & 1.251 & 1.353 & 1.208 & 46.64\% & 1.759 & 1.679 \\
 & & 80 & 0.367 & 0.802 & 0.804 & 0.800 & 14.62\% & 0.992 & 0.986 \\
 & & 100 & 0.396 & 0.728 & 0.735 & 0.732 & 2.71\% & 0.911 & 0.909 \\
 & & 125 & 0.415 & 0.681 & 0.683 & 0.682 & 0.11\% & 0.883 & 0.879 \\
 & & 250 & 0.452 & 0.611 & 0.610 & 0.610 & 0.00\% & & \\
 & & 500 & 0.472 & 0.583 & 0.581 & 0.582 & 0.00\% & & \\
 & & 1,000 & 0.483 & 0.570 & 0.570 & 0.570 & 0.00\% & & \\
\cline{2-10}
 & \multirow{7}{*}{950} & 1,000 & 0.250 & 0.679 & 0.684 & 0.681 & 47.55\% & 0.699 & 0.701 \\
 & & 1,250 & 0.268 & 0.480 & 0.480 & 0.480 & 35.05\% & 0.323 & 0.323 \\
 & & 1,500 & 0.279 & 0.437 & 0.437 & 0.436 & 22.55\% & 0.251 & 0.250 \\
 & & 1,750 & 0.288 & 0.415 & 0.415 & 0.415 & 10.05\% & 0.222 & 0.221 \\
 & & 2,000 & 0.294 & 0.401 & 0.401 & 0.401 & 0.30\% & 0.214 & 0.209 \\
 & & 2,500 & 0.303 & 0.385 & 0.384 & 0.384 & 0.00\% & & \\
 & & 5,000 & 0.321 & 0.359 & 0.359 & 0.359 & 0.00\% & & \\
 \\
\end{tabular}}
\resizebox{0.65\textwidth}{!}{
\begin{tabular}{c|cccc|ccc}
\multicolumn{8}{c}{Panel B: Neurocognitive Aging Data} \\
 & \multicolumn{4}{c}{All Cases} & \multicolumn{3}{c}{Leverage > 1.0} \\
\shortstack{Training Sample\\ \% of Total} & \shortstack{Esti-\\ mated\\ Training\\ Error} & \shortstack{Esti-\\ mated\\ Test\\ Error} & $PRESS/n$ & \shortstack{Pro-\\ posed} & \shortstack{\% of Test\\ Cases} & \shortstack{Esti-\\ mated\\ Test\\ Error} & \shortstack{Pro-\\ posed} \\
\hline
25\% & 0.219 & 0.902 & 0.919 & 0.876 & 30.35\% & 1.895 & 1.798 \\
50\% & 0.284 & 0.466 & 0.467 & 0.467 & 4.56\% & 1.290 & 1.308 \\
75\% & 0.306 & 0.414 & 0.415 & 0.415 & 1.33\% & 1.409 & 1.310 \\

\end{tabular}}
\end{threeparttable}
\caption{Estimated and projected mean squared errors across alternate specifications}
    \label{tab:config-results}
\end{table}

Table \ref{tab:config-results} presents the MSEs overall and for the extreme cases with leverage $>1.0$ as they vary across the different measures and across different specifications for the simulation and application. Each MSE is an average taken across iterations, observations, and levels of model complexity. The results from Table \ref{tab:config-results} indicate that the general patterns observed in Figures \ref{fig:rss} to \ref{fig:hat-error} are not specific to the details of the configurations used and are consistent across the variety of specifications described in Section \ref{Overfitting evaluation}. The proposed estimator continues to be an accurate measure of $MSE^o$. The gap in MSE between the training and test data decreases as the size of the training sample rises, and both $PRESS/n$ and the proposed measure approach the estimated test error as the size of the training sample rises. As can be seen in the Leverage > 1.0 section of the table, the proposed method continues to provide a reasonably accurate forecast of test error, even among these extreme cases with out-of-sample values of $h_i$ that are not possible in the training sample. As the \% of Test Cases shows, for the most overspecified models considered, nearly half the test cases fall into this extreme case of $h_i>1.$

\section{Conclusion} \label{Overfitting conclusion}

This study explores the phenomenon of overfitting in linear regression, whereby models' prediction error is smaller in training samples than in test samples, particularly for complex models. Two key causes are identified for this bias. First, the estimator has access to \emph{forbidden knowledge} in the sense that the in-sample residuals appear in the formula for the estimator, giving it an undue advantage at projecting the outcome variable for cases in the training data. Second, the estimator has \emph{specialized training} that makes it especially effective at explaining the types of movements in predictors that it encountered during fitting.

These sources of overfitting are characterized analytically, and a new estimator is introduced of out-of-sample prediction error in linear regression. The performance of this estimator is assessed in numerical simulations and in an empirical application using data from MRI brain scans. The estimator is found to perform comparably to Predicted Residual Error Sum of Squares (PRESS) for the typical case and to provide more consistently accurate performance across varied test samples. Additionally, and unlike PRESS, the proposed approach can be used to project out-of-sample forecasting error accurately for unusual and customized test samples.


\bibliographystyle{chicago}

\bibliography{overfitting}
\end{document}